\documentclass[aps,prd,twocolumn,showpacs,superscriptaddress,groupedaddress]{revtex4}  
\usepackage{graphicx}  
\usepackage{dcolumn}   
\usepackage{bm}        
\usepackage{amssymb,amsmath}   
\usepackage[english]{babel}
\usepackage[utf8]{inputenc}
\usepackage[T1]{fontenc}
\usepackage{graphicx}
\usepackage{times}
\usepackage{multimedia}
\usepackage{color}
\usepackage[rel]{overpic}
\usepackage{bibentry}
\newcommand{\ep}{\epsilon}

\begin{document}

\title{Particle yields from numerical simulations}

\author{Marietta M. Homor}
\email{homor.marietta.m@gmail.com}
\affiliation{Institute of Physics, E\"otv\"os Lor\'and University, 1/A P\'azm\'any P. S\'et\'any, H-1117 Budapest, Hungary}
\author{Antal Jakov\'ac}
\email{jakovac@caesar.elte.hu}
\affiliation{Institute of Physics, E\"otv\"os Lor\'and University, 1/A P\'azm\'any P. S\'et\'any, H-1117 Budapest, Hungary}

\date{\today}

\begin{abstract}
  In this paper we use numerical simulations to calculate the particle
  yields. We demonstrate that in the model of local particle creation
  the deviation from the pure exponential distribution is natural even in
  equilibrium, and an approximate Tsallis-Pareto-like distribution function
  can be well fitted to the calculated yields, in accordance with the
  experimental observations. We present numerical simulations in
  classical $\Phi^4$ model as well as in the SU(3) quantum Yang-Mills
  theory to clarify this issue.
\end{abstract}


\maketitle

\section{Introduction}

In collider experiments the observed hadron yields are surprisingly
far from the Boltzmann distribution expected from black body radiation
of the hot plasma. It is true for all yields coming from hadron
collisions, starting from p-p collisions in ALICE, CMS, STAR and
PHENIX collaborations, respectively
\cite{Aamodt:2011zj, Khachatryan:2010us, Abelev:2006cs, Adare:2011vy}. 
Part of these particle spectra can be explained by perturbative QCD
calculations \cite{pQCD}, but for a complete description of all of the
fits a Tsallis-Pareto \cite{TSPfit} like Ansatz is necessary.
 These fits have just one additional parameter compared to a Boltzmannian, yet they work very nicely in experimental fits
\cite{Khachatryan:2010us, Khachatryan:2010xs, Barnafoldi:2011zz, Cleymans:2011in}
although in certain cases the soft and hard physics has to be treated
separately \cite{Urmossy:2015kva}. 


There are several interpretation of these results. A QCD-based
generalized Ansatz
\cite{Hagedorn1, Wong:2014uda, Wong:2015mba} results in
distribution functions that are similar to Tsallis distribution. More
natural explanation is to assume some collective behaviour that leads
to these type of distributions. The reason for the deviation from the
Boltzmann distribution can be the finite volume \cite{Biro:2012bka},
fluctuating temperature \cite{Wilk:2012zn, Biro:2014fna}. In fact
the leading deviation from Boltzmann distribution is of Tsallis
form \cite{GyorgyiG}.  It is also possible that the event-by-event
distribitions are Boltzmannian, but the hadron multiplicities
fluctuate according to negative binomial distribution, and only in the
cumulative yields do we see Tsallis distribution
\cite{Urmossy:2015kva}.

In this paper we suggest another possible mechanism to observe
non-Boltzmann distribution. It is possible, namely, that the energy
levels of the system are Boltzmann distributed, still the observed
particle yields follow a different distribution. To understand this
possibility, we have to rethink the mechanism, how particles emerge
from a strongly interacting plasma.

In weakly interacting gases, like the photon gas that interacts only
with the wall of the cavity, the particle states are (almost) the same
as the energy eigenstates: they have definite momentum and energy, and
they extend to the whole cavity. If a photon escapes from the system
it carries information about the distribution of the energy
eigenstates, and so, correspondingly, we obtain a photon yield that is
distributed according to the Bose-Einstein distribution.

In strongly interacting plasmas, however, the situation
changes. Particle states are no longer energy eigenstates, in fact
they consist of a lot of energy eigenstates with a Lorentzian
envelope: they are quasiparticles. Moreover, they are basically local
objects, they do not extend to the whole plasma. The phenomenon of jet
quenching \cite{Wiedemann:2009sh} clearly indicates that high energy
particles are created locally, in a volume of at most of order one
fm. As a consequence, if they escape the plasma, they do not carry
information about the occupation of the global energy levels, but
about the local energy density. The fluctuation of the energy in a
small volume, however, is different than in a large subsystem, it need
not (and actually do not) follow Boltzmann distribution.

According to this picture to assess the particle yields coming from a
strongly interacting plasma we have to measure the \emph{distribution
  of the local energy density}. This is a measurable quantity even in
numerical simulations, making possible to give predictions on the
observable yields which are otherwise very hardly accessible quantities.

In this paper we have considered two models basically to demonstrate the
method: a classical $\Phi^4$ model and a quantum SU(3) Yang-Mills
gauge theory. As it turns out, in both cases the distribution of the
local energy density stabilizes relatively fast, well before the
actual thermal equilibration, and a Tsallis distribution is an
excellent fit to them. In the process of approaching equilibrium, the
parameters of the Tsallis distribution change. In classical scalar
model case it is possible to follow real time evolution, and so the
variation of the temperature and Tsallis parameter in real time. In
the followings we discuss the histogram method to determine the local
energy distribution function, consider the classical quartic model and
the quantum Yang-Mills model, and finally we close the paper with a
Conclusion section.

\section{Local energy density distribution}
As stated in the introduction, our aim is to determine the local
energy density distribution. In this section, we recall some basic
definitions of probability theory.

First of all, let $X$ be a stochastic variable. The indicator of $X$
being in the \mbox{$[x,x+\Delta x]$} interval is
$\mathbb{I}_{[x,x+\Delta x]}(X) = \Theta(X-x)\Theta(x+dx-X)$, where
$\Theta$ is the Heaviside function. The expectation value of this
indicator equals the probability of $X$ being in the interval:
\begin{equation}
  \langle \mathbb{I}_{[x,x+\Delta x]}(X)\rangle = \mathcal{P}(X\in[x,x+\Delta x]).
  \label{eq:indicator}
\end{equation}

In statistical approach we take the expectation value above some
configuration space, then we assume that $X$ is a function(al) of the
configurations $X(A)$. If the distribution of the configurations $A$
is given, let us denote it by $f(A)$, then the expectation value of a
general quantity $R(A)$ can be computed as
\begin{equation}
  \label{eq:microscopicf}
  \langle R \rangle = \int\!{\cal D}\!A\, R(A) f(A).
\end{equation}
In canonical ensemble $f(A)=\exp(-\beta H(A))/Z$. In practice,
however, we generate a lot of configurations according to the
distribution function $f(A)$, either by solving the equation of motion
(classical theory), by following a Markov-process (Monte Carlo
simulations) or considering small subsystems of a configuration in a
large volume. In any case we have $A_1,\,A_2,\dots \,A_n$
configurations and we take the expectation value by summing above
them:
\begin{equation}
  \langle R \rangle = \frac1n \sum_{i=1}^n R(A_i).
\end{equation}
Therefore the expectation value of the indicator is proportional to
the number of configurations $n_i$, where the $X(A_i)$ quantity have
value between $x$ and $x+\Delta x$. Therefore
\begin{equation}
  \mathcal{P}(X\in[x,x+\Delta x]) = \frac{n_i}{n}.
\end{equation}
To determine the complete distribution function, therefore, we divide
the possible outputs of $X$ to bins, each of them is $\Delta x$
wide. Then we scan over all available configurations, and each
configuration contributes to the bin that contains $X(A_i)$. This
provides a histogram that is exactly the desired
$\mathcal{P}(X\in[x,x+\Delta x])$ distribution function.

The limit $\Delta x\to 0$ provides the probability density of the
variable $X$:
\begin{equation}
  p(x)=\lim_{\Delta x \rightarrow 0+} \frac{\mathcal{P}(X \in
    [x,x+\Delta x])}{\Delta x}. 
  \label{eq:pdf}
\end{equation}
If we write (\ref{eq:indicator}) into the definition (\ref{eq:pdf}) of
$f(x)$, then we get a Dirac-$\delta$ approximation and we can write
rather formally:
\begin{equation}
  p(x)=\langle \delta ( X -x) \rangle.
  \label{eq:deltadens}
\end{equation}
In this work, the quantity in question -- in other words the
stochastic variable -- is the local energy density $\epsilon_x$ where
$x$ is an arbitrary space-time coordinate. With this
(\ref{eq:deltadens}) becomes:
\begin{equation}
  p(\epsilon)=\langle \delta (\epsilon_x - \epsilon) \rangle.
  \label{eq:endens}
\end{equation}

It is important to note that the local energy density, or the energy
in a small volume need not to follow Boltzmann distribution, even if we
use canonical ensemble to generate configurations. The reason is that
in a small volume the energy contains a considerable amount of surface
energy, too. The surface energy depends both on the state of the
singled out volume and on its environment, and so we cannot associate it
to any of them. Only in largish volumes which are small compared to the
complete system, but large enough to neglect the surface energy terms,
can we deduce that the probability density of measuring a given energy
value is Boltzmann-distributed.

\subsection{Local energy distribution in free systems}

It is worth to think about the form of the local energy distribution
when it is Boltzmann-like. So let us assume that $p(\epsilon)={\cal N}
e^{-c\epsilon}$ with some constants. We have two constraints:
\begin{equation}
  1 = \int\limits_0^\infty d\epsilon\, p(\epsilon),\qquad
     \langle \epsilon_x \rangle = \int\limits_0^\infty
     d\epsilon\,\epsilon p(\epsilon),
\end{equation}
these fix the constants to be
\begin{equation}
  p(\epsilon) = \frac1{\langle \epsilon_x \rangle}
  e^{-\epsilon/\langle \epsilon_x \rangle}.
\end{equation}
Therefore we do not expect $e^{-\beta \ep}$ form, only if $\langle
\epsilon_x \rangle=T$.

In simple systems we in fact obtain such a form. Most simply, in a
system built up from local independent systems we have for all
configurations $\sigma$
\begin{equation}
  E(\sigma) = \sum_i \ep_i(\sigma_i).
\end{equation}
In this case 
\begin{equation}
  p(\ep)=\frac1Z \sum_{\sigma} e^{-\beta E}\delta(\ep-\ep_i) =
  \frac1{Z_i} \sum_{\sigma_i} e^{-\beta \ep_i}\delta(\ep-\ep_i) =
  \frac1Z e^{-\beta\ep},
\end{equation}
because at sites $j\neq i$ the corresponding $Z_j$ factors drop out.

We also have the same results in case of free systems, even when the energy is the sum of the momentum states, while the local
energy density is localized in real space. To prove this statement we
first rewrite the energy density distribution as
\begin{equation}
  p(\epsilon) = \Theta(\epsilon)\! \int\limits_{-\infty}^\infty \!
  d\lambda \, \left\langle e^{i\lambda (\ep-\ep_x)}\right\rangle 
  = \Theta(\epsilon)\! \int\limits_{-\infty}^\infty \!
  d\lambda \,e^{i\lambda \ep} \left\langle e^{-i\lambda\ep_x}\right\rangle,
\end{equation}
where we assumed $\ep_x>0$ for all configurations. Then we expand the
exponential
\begin{equation}
  \left\langle e^{-i\lambda\ep_x}\right\rangle = \sum_{\ell=0}^\infty
  \frac{(-i\lambda)^\ell}{\ell!} \left\langle \ep_x^\ell\right\rangle.
\end{equation}
To avoid UV divergences we renormalize the above expression taking the
normal ordered product, so we calculate
\begin{equation}
  \left\langle : \ep_x^\ell :\right\rangle.
\end{equation}
The local energy density can be defined in a number of ways, each
definitions differ from each other in total divergences. We will
choose a simple representation, where the local energy density can be
written as function of the creation-annihilation operators $a_p$ and
$a^\dagger_p$ in  $d$ dimensions as
\begin{equation}
  :\ep_x: = \int \frac{d^dp}{(2\pi)^d}\frac{d^dq}{(2\pi)^d} \sqrt{\omega_p\omega_q}
  e^{i(p-q)x} a_q^\dagger a_p.
\end{equation}
It is simple to see that 
\begin{equation}
  :H:=\int d^dx :\ep_x: =  \int\frac{d^dp}{(2\pi)^d} \omega_p a_p^\dagger a_p.
\end{equation}
Now we can compute the expectation value in question at \mbox{$x=0$}:
\begin{eqnarray}
   \left\langle : \ep_{x=0}^\ell :\right\rangle 
  =&& \frac1Z\sum_{\{n\}} e^{-\beta E_n}\int 
   \prod_{i=1}^\ell \frac{d^dp_i}{(2\pi)^d}\frac{d^dq_i}{(2\pi)^d}
   \sqrt{\omega_{p_i}\omega_{q_i}}\times\nonumber\\
  &&\times \left\langle n\left| a^\dagger_{p_1}\dots
       a^\dagger_{p_\ell} a_{p_1}\dots a_{p_\ell}\right|
     n\right\rangle,
\end{eqnarray}
where $E_n=\langle n| :H:|n\rangle$. Since the same state stands in the left and
right hand side of the expectation value,
 we must have the same number of creation and annihilation operators
for each momenta. We will omit the possibility that more than two
operators have the same momenta, since these contributions are
suppressed by factors of $V_d$ the volume of the $d$ dimensional
space. This means that we have to make pairs (Wick theorem). It is
easy to see that all pairings give the same contribution, so finally
we have
\begin{equation}
  \left\langle n\left| a^\dagger_{p_1}\dots
      a^\dagger_{p_\ell} a_{p_1}\dots a_{p_\ell}\right|
    n\right\rangle = \ell! \prod_{i=1}^\ell n_{p_i} (2\pi)^d\delta(p_i-q_i).
\end{equation}
Substituting back this result we see that the expectation value of the
$\ell$-times local energy density is proportional to the expectation
value of the local energy density to the $\ell$th power:
\begin{equation}
  \left\langle : \ep_{x=0}^\ell :\right\rangle = \ell! \left\langle :
    \ep_{x=0} :\right\rangle ^\ell.
    \label{eq:exppowexp}
\end{equation}
Therefore
\begin{equation}
  \left\langle e^{-i\lambda\ep_x}\right\rangle = \sum_{\ell=0}^\infty
  (-i\lambda)^\ell\left\langle \ep_x\right\rangle^\ell =
  \frac1{1+i\lambda \left\langle \ep_x\right\rangle},
\end{equation}
and so the inverse Fourier transform yields:
\begin{equation}
  p(\ep) =\frac1{\left\langle \ep_x\right\rangle} e^{-\ep/\left\langle
      \ep_x\right\rangle},
\end{equation}
which means that in the free systems the local energy density is
indeed Boltzmann-distributed.

We see from this calculation that the validity of the Boltzmann
distribution depends on very sensitive details, for example that the
expectation value of powers of the local energy density is
proportional to powers of the expectation value of the local energy
density (cf. eq.\ref{eq:exppowexp}). In a general theory it will not be true anymore, resulting
that $\left\langle e^{-i\lambda\ep_x}\right\rangle$ is not a simple
pole and then $p(\ep)$ is no longer exponential. We can not determine
the actual form, but based on very general arguments \cite{GyorgyiG}
we expect that if the deviation is small, then it must be a
Tsallis-Pareto distribution.

In the following sections, we consider the real time simulation of the
classical $\Phi^4$ theory in 3 dimensions and perform a standard Monte
Carlo simulation with heat-bath algorithm for the Euclidean SU(3)
gauge theory.

\section{A toy model: classical $\Phi^4$ theory}

Our first toy model is the well-known classical $\Phi^4$ theory.  One
of the advantages of classical theories is that we can perform
real-time simulations by successively solving the canonical equations
and we can calculate physical quantities that are hardly accessible in
other methods. Classical theories are used to approach the full theory
in a lot of contexts 
\cite{Aarts:2000mg,
Borsanyi:2003ib,
Destri:2004ck,
Sexty:2005pz,
Romatschke:2006nk,
Berges:2011sb,
Jin:2012qh,
Berges:2014yta,
Berges:2015ixa,
Homor:2015qza}.

The discretized version has the Hamiltonian \cite{Homor:2015qza}
\begin{equation}
  H = \sum_{x\in U} \epsilon_x,
\end{equation}
where $U$ denotes the discretization mesh (in our case a cubic lattice
with $N$ sites in all directions, $N=40,\,50$), and $\epsilon_x$ is the
local energy density
\begin{equation}
\epsilon_\mathbf{x}=\frac{1}{2}\Pi_{\mathbf{x}}^2 + \frac{1}{2}(\nabla
\Phi)_{\mathbf{x}}^2 + \frac{m^2}{2}\Phi_{\mathbf{x}}^2 +
\frac{\lambda}{24}\Phi_{\mathbf{x}}^4. 
\end{equation}
In this expression we have to use the discretized gradient
$\nabla_i\Phi(x) = a^{-1}[\Phi(x+a e_i)-\Phi(x)]$, where $a$ is the
discretization spacing, and $e_i$ is the unit vector pointing to the
$i$th direction. The corresponding equations of motion read
\begin{equation}
  \dot \Phi =\Pi,\qquad \dot\Pi=\triangle\Phi -
  m^2\Phi-\frac\lambda6\Phi^3, 
\end{equation}
where
$\triangle\Phi=a^{-2}\sum_{i=1}^3[\Phi(x+ae_i)+\Phi(x-ae_i)-2\Phi(x)]$
is the discretized Laplacian. The continuous equation of motion
preserves energy, but in the time discretized version the energy
conservation depends on the algorithm. We used leapfrog and
Runge-Kutta methods; for further discussion cf. \cite{Homor:2015qza}.

We note that the system can be rescaled as
$t\to t/a,\, \Phi\to \sqrt{\lambda}\,a\Phi,\, \Pi\to
\sqrt{\lambda}\,a^2\Pi$,
then we have the same equations of motion with $a=1$, $ m\to am$ and
$\lambda=1$. This means that the value of $\lambda$ does not modify
the classical dynamics, it can be compensated by the normalization of
the fields. The energy density rescales as
$\epsilon\to \lambda a^4 \epsilon$.  In the simulations we have used
the $a=1$ unit, but we have kept the value of $\lambda$ to test the
numerical effects.

After thermalization we can use the thermodynamical notions. The
temperature ($T$) of the system is defined as \cite{Homor:2015qza}:
\begin{equation}
T=\frac{1}{2 N^3}\langle\vert\Pi_{k}\vert^2\rangle,
\end{equation}
where $N^3$ is the number of lattice sites and $\Pi_{k}$ is the
Fourier-transformed momentum field. We use this formula to check
whether the system reached thermal equilibrium by verifying that
$\langle\vert\Pi_{k}\vert^2\rangle$ is independent of $\mathbf{k}$
(equipartition). It turned out, that we can distinguish two time
scales, as higher modes thermalise much faster than low ones. After
$10000$ time steps, the system can be considered fully thermalised.

\subsection{Numerical results for the classical $\Phi^4$ theory}
\begin{figure}[h!t]
\centering
\begin{overpic}[scale=0.68]{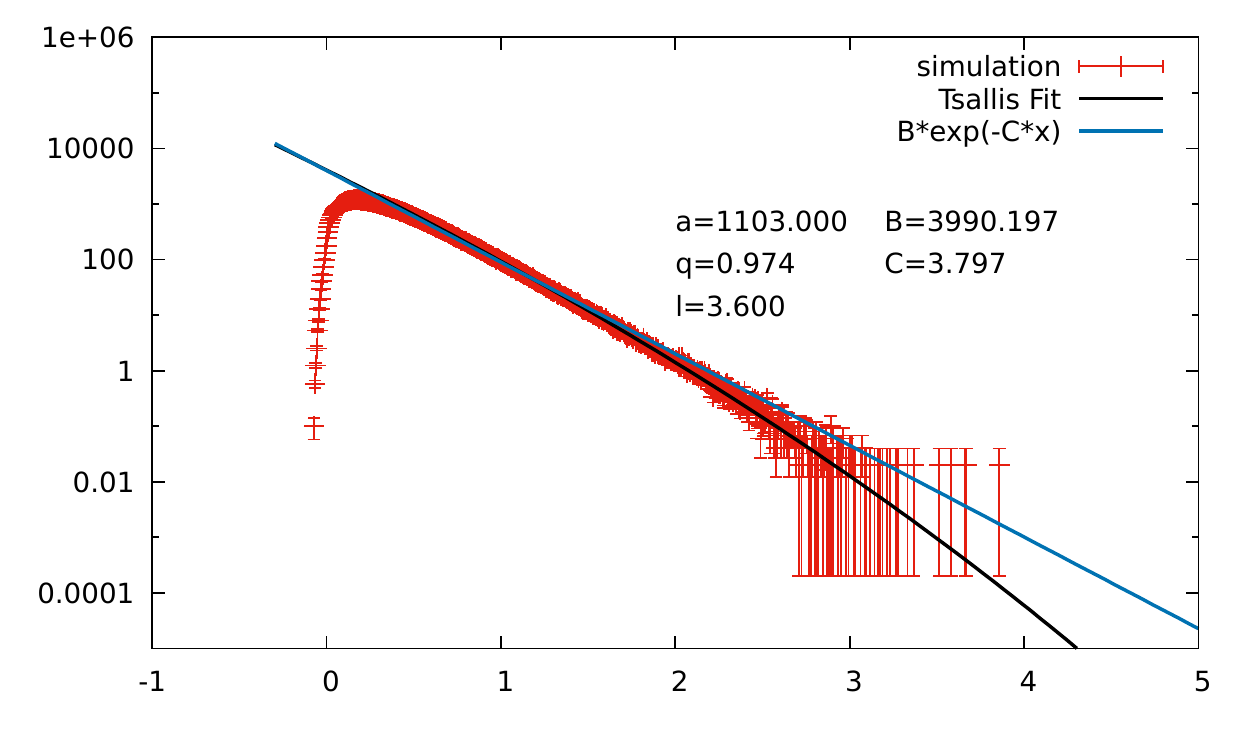}
 	\put (-1,20){\rotatebox{90}{{\small Num. of data (log)}}}
    \put (45,-1){{\small $\epsilon_x$ (arb. unit)}}
\end{overpic}
\caption{Local energy density histogram with Boltzmann (blue line) and Tsallis (black line) fits on semi-logscale after $17$ time steps. Simulation with random initial condition $\Pi_\mathbf{x}$. Data points are averaged from $50$ simulations and shown with their standard error.}
\label{fig:logenrandom}
\end{figure}

\begin{figure}[h!t]
\centering
\begin{overpic}[scale=0.68]{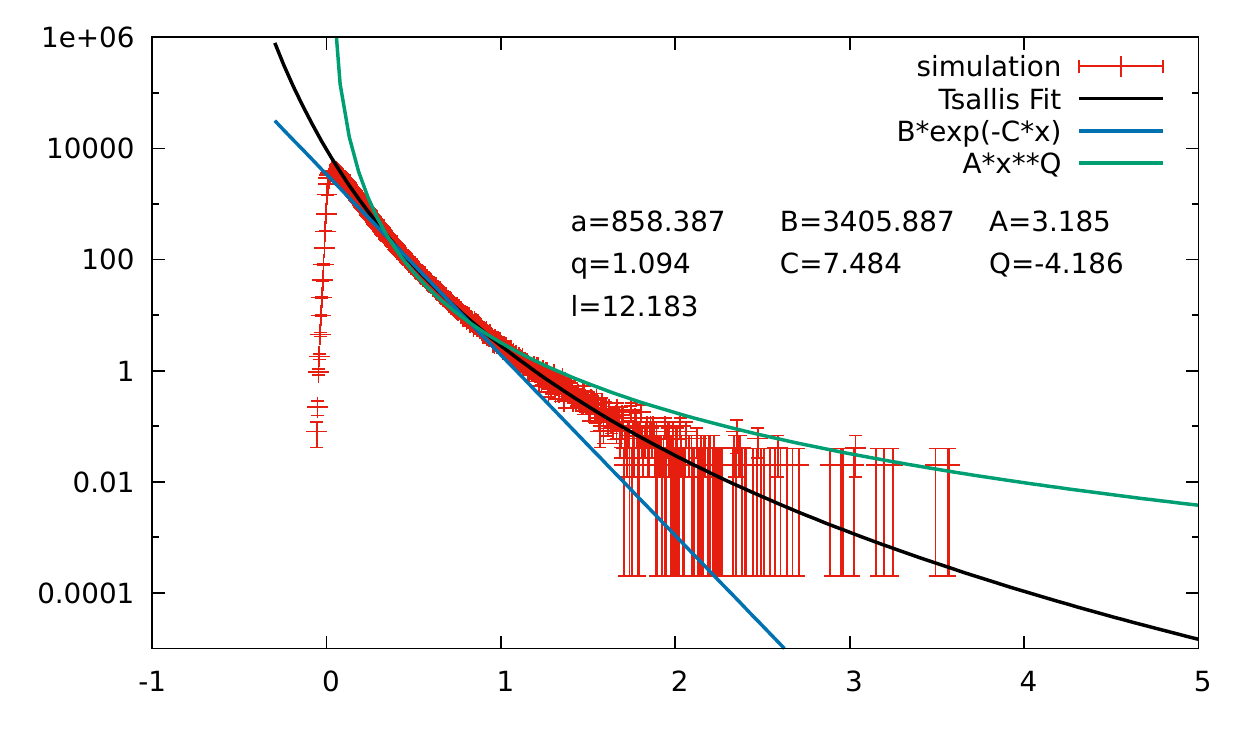}
 	\put (-1,20){\rotatebox{90}{{\small Num. of data (log)}}}
    \put (45,-1){{\small $\epsilon_x$ (arb. unit)}}
\end{overpic}
\caption{Local energy density histogram with simple power function (green line), Boltzmann (blue line) and Tsallis (black line) fits on a semi-logscale after $17$ time steps. Simulation with secant hyperbolic initial condition for $\Pi_\mathbf{x}$. Data points are averaged from $50$ simulations and shown with their standard error.}
\label{fig:logensech}
\end{figure}

We have solved the classical EoM on $40^3$ and $50^3$ lattices. As
initial conditions we have chosen $\Phi_x=0$ for every point, and we have
assumed pointwise independent distributions for the canonical
momenta. In one case the momentum distribution was a uniform
distribution in the $[0,1]$ range, in the other case we had a
$1/\cosh(x)$ distribution. In the simulation we have chosen $dt=0.1$
time step in lattice spacing units.

After starting the simulation, the distribution of the local energy
density very quickly stabilizes. Already after the 17th time step the
histograms reach the characteristic form which remained true in all
later times. We can see these distributions in figures
\ref{fig:logenrandom} and \ref{fig:logensech}. Each data point in the
histogram is the average of $50$ runs with the same initial condition
and the errorbars represent the standard error of the mean (SEM). It
is apparent that the distribution deviates from the Boltzmannian
exponential form, but a \emph{Tsallis-Pareto distribution} proved to
be a very good fit
\begin{equation}
  \mathcal{P}(\varepsilon)=a \left[ 1+ (q-1)\beta \varepsilon
  \right]^{\frac{1}{1-q}}.
\label{eq:classicTs}
\end{equation}
Note that for $q \rightarrow 1$ it gives back the
Boltzmann distribution. The actual values, $q=0.974$ and $q=1.094$
respectively, are very close to the Boltzmannian case, and so it can
be revealed only by a thorough analysis with at least $10^6$ independent
data points.

\begin{figure}[h!t]
\centering
 \begin{overpic}[scale=0.68]{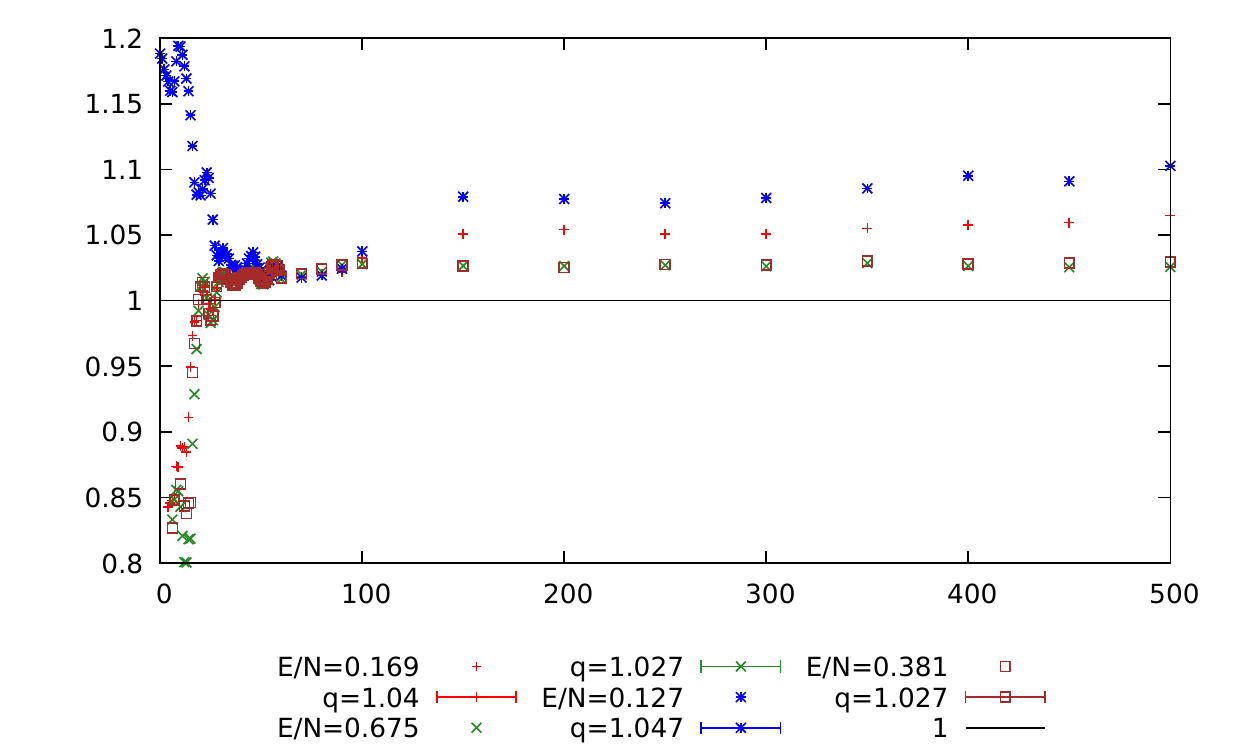}
   \put (3,25) {\rotatebox{90}{{\small Tsallis $q$ parameter}}}
    \put (43,9){{\scriptsize Num. of time steps}}
   \end{overpic}
  \caption{Time dependence of the Tsallis parameter at 4 different total energies with various initial conditions. The first $500$ time steps.}
  \label{fig:Tsallismidlong}
\end{figure}

\begin{figure}[h!t]
\centering
 \begin{overpic}[scale=0.68]{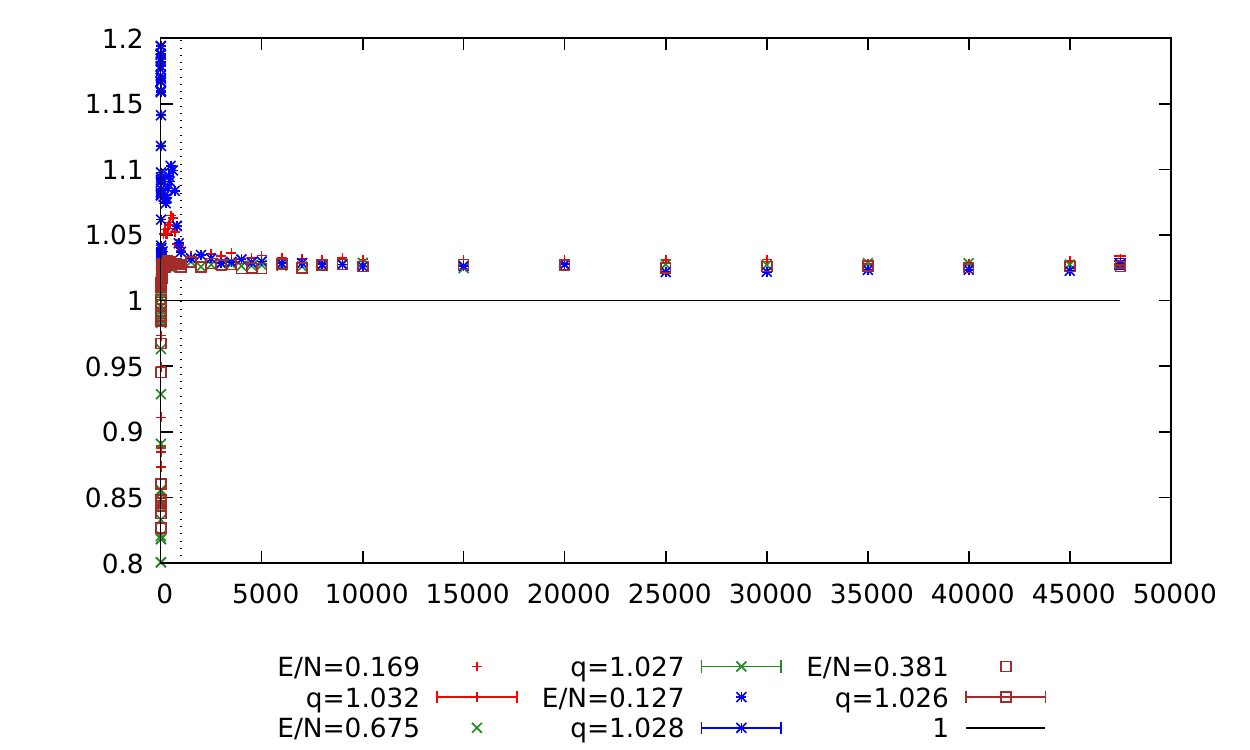}
   \put (3,25) {\rotatebox{90}{{\small Tsallis $q$ parameter}}}
    \put (43,9){{\scriptsize Num. of time steps}}
   \end{overpic}
  \caption{Time dependence of the Tsallis parameter at 4 different total energies with various initial conditions. The last points indicate the average values for each energies from $1000$ to $45000$ time steps.}
  \label{fig:Tsallislong}
\end{figure}

In the 17th time step the system is very far from equilibrium, but the
Tsallis-Pareto distribution of the energy density remained true, with
time dependent Tsallis parameter $q(t)$. This function is plotted in
Figures \ref{fig:Tsallismidlong} and \ref{fig:Tsallislong} for two
different time intervals, starting with different total energy
(corersponding to different temperatures after thermalization and
different lattice sizes). We can observe that even in that cases when
$q$ started from a value smaller than 1, finally in all runs it
reached a value that is consistently larger than 1. It is interesting
that this value seems to be independent on the temperature as well as
on the lattice sizes we studied. The Tsallis-parameter takes its
equilibrium value already in the pre-thermalised state (i.e. when only
higher modes are thermalised) within error. The actual value of the
equilibrium Tsallis parameter is $q=1.024$ is in the order of the
experimental values.

At this stage it is not clear, whether the deviation from the
Boltzmann distribution is a property of the energy density only, or
the microscopic distribution function $f(A)$ over the configurations
(cf. \eqref{eq:microscopicf}) is also non-Boltzmannian. To this end we
have carefully studied the distribution of $\Pi_x^2$ for different
energy values and different lattice sizes with large statistics. If
the thermal ensemble has a distribution function $e^{-\beta H}$, then
the $x=\Pi_x^2/2$ values for a given $x$ must follow $e^{-\beta x}$
expontential distribution, i.e. we must have $q=1$ in the Tsallis fit.
The result for the averaged data is
\mbox{$q_{momentum} = 0.999 \pm 0.001$} for all the cases we
studied. This means that the distribution over the configurations is
in fact the standard Boltzmann distribution, and only the energy
density has a Tsallis-like distribution. Learning this fact we could
proceed to the quantum field theory case, where only the equilibrium
can be studied with Monte Carlo methods, but the distribution of the
energy density is expected to be Tsallis-like even in this case.

\section{Euclidean SU(3) pure gauge theory}

Our second model is the quantum SU(3) Yang-Mills theory. The action of
the model in Euclidean formalism is the following:
\begin{equation}
S_{YM}=\frac{1}{4}\int d^4x \,F^a_{\mu \nu}F^a_{\mu \nu},
\end{equation}
where $F_{\mu \nu}(x)=-\mathfrak{i}gF^a_{\mu\nu}(x)T_a$ is the gluon field strength tensor, $T_a$ are the generators of the Lie-algebra and $g$ is the coupling constant.

We use the Wilson-action for the lattice formulation of the theory:
\begin{equation}
S[U]=\sum_p \beta\left( 1-\frac{1}{N}\mathrm{Re}\, \mathrm{Tr}\, U_p\right),
\end{equation}
where $U(x+\mu ,x)$ is the parallel transporter from lattice
coordinate $x$ to $x+\mu$ and the plaquette variable corresponding to
$x$ is the product of four parallel transporters along the closed path
$x\rightarrow x+\mu \rightarrow x+\mu +\nu \rightarrow x+\nu
\rightarrow x$.

It is well-known that the Wilson-action corresponds to the continuum
theory if one chooses \mbox{$\beta = 2 N / g^2$} (in our case $N=3$)
and the connection between the parallel transporters and the gauge
fields is \mbox{$U(x,\mu)=\textrm{e}^{-a A_\mu(x)}$}. Continuum limit
is reached as $\beta\to\infty$.

The local energy density is now given by the plaquette energy:
\begin{equation}
\epsilon=\left\langle 1 - \frac{1}{\mathrm{Tr}\mathbb{I}}\mathrm{Tr}U_p\right\rangle.
\end{equation}
We determine the histogram of the local energy density in the same way
as we have done in the classical theory, picking out independent
configurations from the thermal ensemble.

An advantage of the histogram method is that renormalization can be
explicitly traced in the distributions. In case of the local energy
density, being a composite operator, we expect a multiplicative
renormalization as well as an eventual mixing with the unit operator
(additive renormalization). In the histograms the two types of
renormalization show up as a dilatation and a position shift. Neither
of these effects modify the power of the high-energy tail: therefore
the Tsallis parameter is not renormalized.

We use Monte Carlo simulation with the well-known heat-bath algorithm
to determine the distribution of $\epsilon$ with zero energy initial
condition (i.e. \mbox{$\epsilon_p=0$} for all plaquettes) at lattices $N_t\times N_s^3$.

For the fits of the resulting distributions we have taken into account
the reduction of the phase space at low energy densities.  As a
result, we modify our previous fit function (\ref{eq:classicTs}) by a
factor of $x^n$ where $n\approx3$:
\begin{equation}
f(x)=ax^n (1+(q-1)\,l\,x)^{\frac{1}{1-q}}.
\label{eq:su3Ts}
\end{equation}
The error of the histogram is assumed to be a Gaussian (meaning
$\sqrt{N}$ standard deviation for a bin which contains $N$
points). However, in more realistic models it would be advised to
perform a maximum likelihood parameter estimation (in the context of
Tsallis distributions cf. \cite{Shalizi:2017}).

\subsection{Plaquette energy histogram}
\begin{figure}[h!t]
\centering
 \begin{overpic}[scale=0.68]{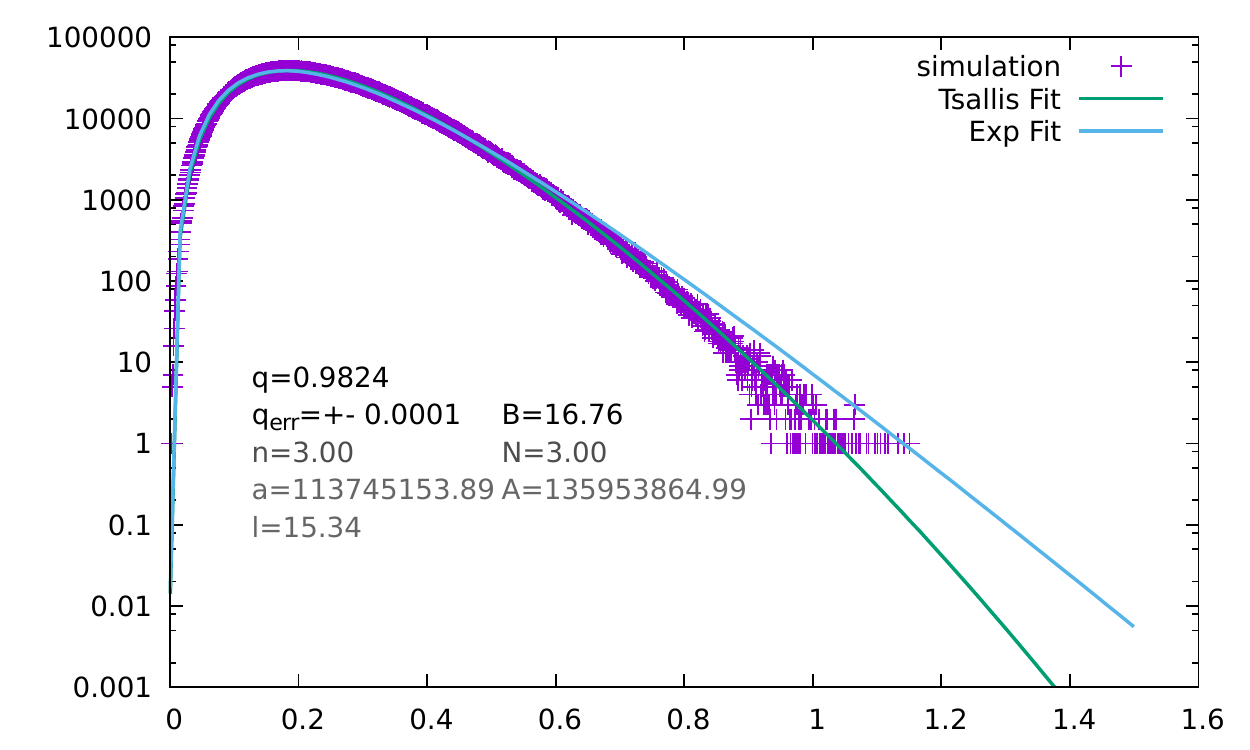}
   \put (1,20) {\rotatebox{90}{{\small Num. of data (log)}}}
    \put (43,-2){{\small Plaquette energy $\epsilon_x$}}
   \end{overpic}
  \caption{Plaquette energy histogram on semi-logscale after $15$ heat-bath sweeps at \mbox{$\beta=8$}, \mbox{$ N_t=8$}, \mbox{$ N_s=60$}. The blue line is the Boltzmann and the green line is the Tsallis fit.}
  \label{fig:plaqenhist}
\end{figure}

A typical plaquette energy histogram is presented in
Fig.~\ref{fig:plaqenhist}. In the figure the best {Tsallis
  (\ref{eq:su3Ts})} and Boltzmannian
\mbox{$g(x)=Ax^N \textrm{e}^{-B x}$} fits can also be seen. It is
evident that the Tsallis fit is better than the Boltzmannian one,
similarly as in the case of classical $\Phi^4$ theory.

\subsection{Tsallis $q$ for various $\beta$}

\begin{figure}[!ht]
\centering
\begin{overpic}[scale=0.68]{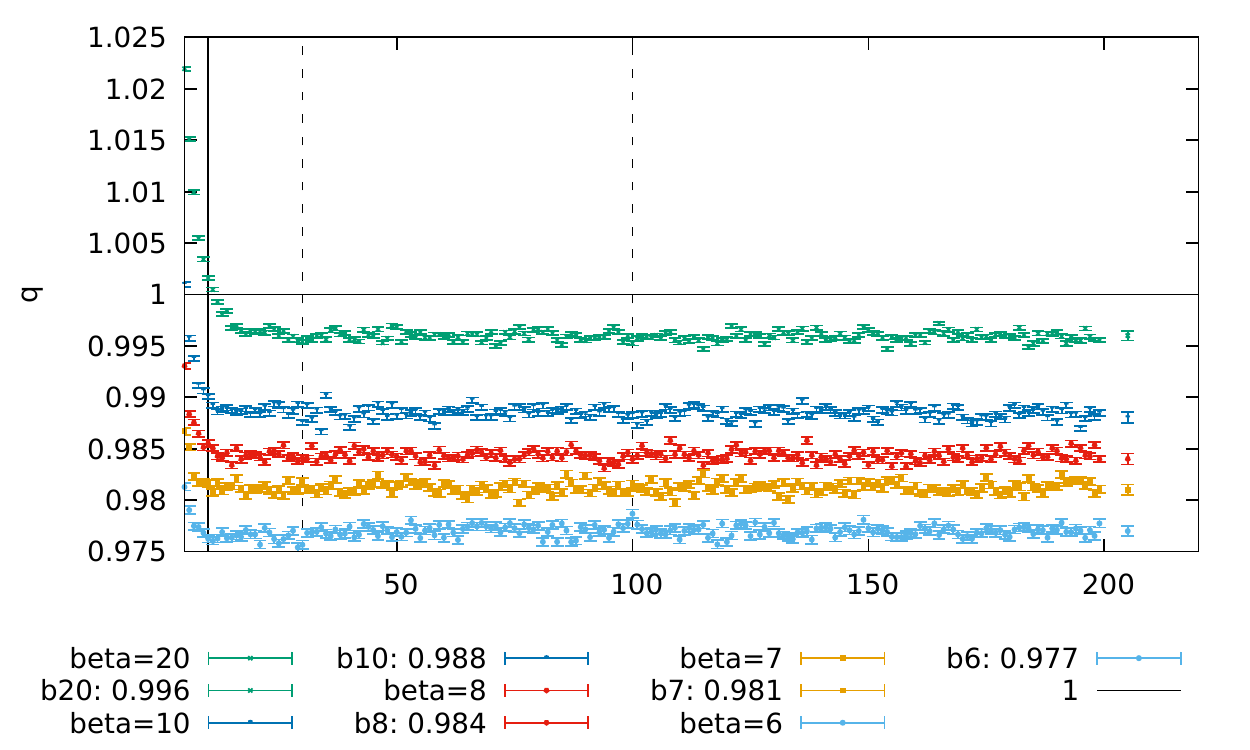}
    \put (40,10){{\tiny Num. of MC time steps $\tau$}}
   \end{overpic}
   \caption{Tsallis $q$ parameter during MC sweeps for various \mbox{$\beta$}, with \mbox{$N_t=2$} and \mbox{$N_s=50$}. The last, slightly separated points show the average from \mbox{$\tau=100$} with their statistical error. The average values are present in the legend.}
   \label{fig:qvarbeta}
\end{figure}

The evolution of the Tsallis parameter (with its fit error) is
presented in Fig. \ref{fig:qvarbeta} as a function of the Monte
Carlo time. We covered a wide range of $\beta$, starting from
$\beta=6$ to as high values as $\beta=20$. Note, that the plaquette
energy has an upper bound (due to the properties of the trace of SU(3)
matrices), and this severely distorts the histogram below
$\beta=6$. 

It is interesting that the shape of the distribution shows up very early,
well before thermalization: already after the first \mbox{MC sweep} we
find a roughly Tsallis-like distribution, although the Tsallis fit
converges well after about $\tau=10$ (first vertical line in
Fig. \ref{fig:qvarbeta}). Thermalization time is $\tau\approx30$
(second vertical line in Fig. \ref{fig:qvarbeta}). We have calculated
the equilibrium Tsallis parameter from configurations after $\tau=100$
(third vertical line) for each simulation.

\subsection{Tsallis $q$ for various $N_t$}
To connect the numerical observations to physical units, we have to
fix the scale. We have chosen the Sommer-scale \cite{Sommer:1993ce} and
interpolated the $\beta$ and lattice constant $a$ relation based on
the data from \cite{Francis:2015lha, Necco:2001xg, Guagnelli:1998ud}. The dimensionless temperature is as follows:
\mbox{$r_0T=\frac{1}{N_t}\frac{r_0}{a}$}, where \mbox{$r_0\approx0.5$
  fm} is the Sommer-scale parameter.

\begin{figure}[!ht]
\centering
\begin{overpic}[scale=0.68]{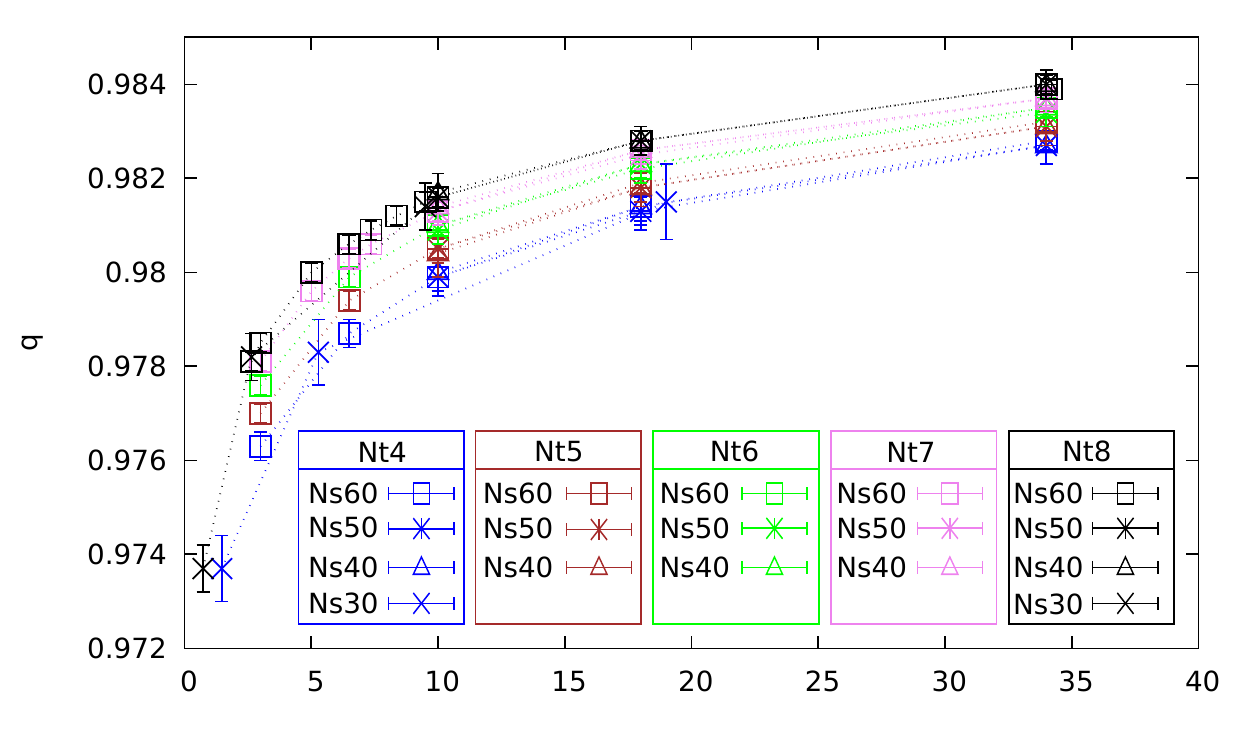}
    \put (40,0){{\small Temperature ($r_0 T$)}}
   \end{overpic}
   \caption{Temperature dependence of $q$ for various $N_t$ and $N_s$. Dashed lines are to guide the eye.}
   \label{fig:qvsT}
\end{figure}

The temperature dependence of the $q$ parameter is shown in Fig.
\ref{fig:qvsT}. Five different $N_t$ value is considered. To check the
thermodynamic limit, we repeated the simulation for \mbox{$N_s=60$},
\mbox{$N_s=50$} and \mbox{$N_s=40$}.  

\subsection{Continuum limit}

\begin{figure}[!ht]
\centering
\begin{overpic}[scale=0.68]{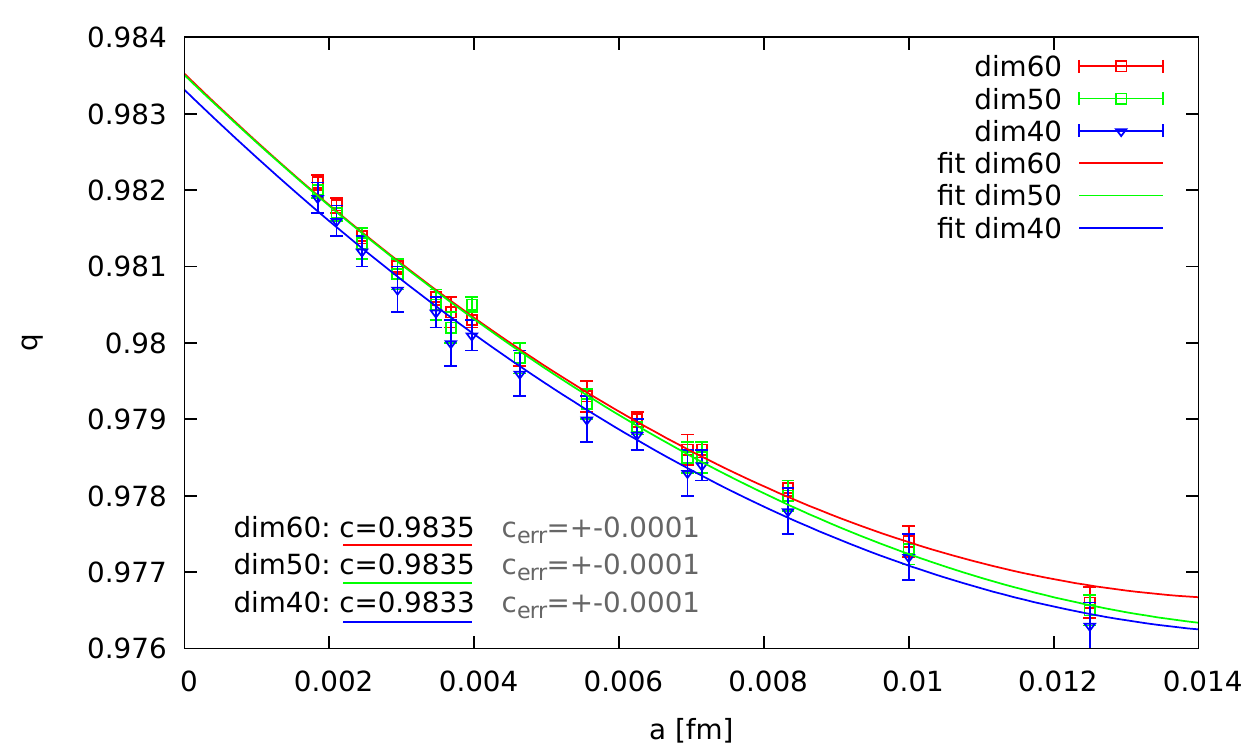}
    \put (73,37){{\footnotesize $\mathrm{A} x + \mathrm{B} x^2+\mathrm{C}$}}
   \end{overpic}
   \caption{Tsallis $q$ parameter against the lattice constant. Second order polinomial fit is performed to the numeric data. Three different lattice size is taken into consideration. \mbox{Red: $N_s=60$}, \mbox{green: $N_s=50$} and \mbox{blue: $N_s=40$}} 
   \label{fig:ContLim}
\end{figure}

At $T=0$, collecting all the measured values at different lattice
spacing, we can determine the continuum limit of the Tsallis
parameter. Our result is presented in Fig. \ref{fig:ContLim}. We
used second order polinomial to fit the numeric data and aquired
$q=0.9835\pm0.0005$ for \mbox{$N_s=60$} and \mbox{$N_s=50$} as
well. Interestingly enough, we got \mbox{$q<1$} as opposed to the case
of the classical $\Phi^4$ theory, though the absolute difference from
$1$ is approximately the same.  

\section{Conclusions and outlook}

The particles emerging from a strongly interacting plasma are created
locally and so they carry information about the local energy
density. The distribution of this quantity is in general different
from the canonical energy level distribution, except for the free (or
very weakly interacting) theories. Therefore in the particle yields
coming from a strongly interacting plasma we should not expect
Boltzmann distribution. For the actual expectation we have to
measure the distribution of the local energy density.

In this work we have determined the local energy density distribution
with histogram method in the classical $\Phi^4$ and in the quantum
SU(3) Yang-Mills theory. In both cases the energy level distribution
is Boltzmannian, but we have found that the Boltzmann distribution
does not fit well to the local energy distribution in either
case. However, the Tsallis distribution is a good fit, similarly to
experimental data. The corresponding Tsallis parameter differs
significantly from 1. Thermodynamic limit analysis is performed in
both cases and we carried out the continuum limit analysis as well for
SU(3) gauge theory. We remark that the renormalization of the local
energy density does not affect the power of the power law tail,
i.e. the Tsallis parameter $q$ receives no renormalization correction.

We have found that in case of the classical $\Phi^4$ theory, the
Tsallis parameter $q=1.024\pm0.001$ -- this is in the order of the
experimental values obtained from heavy ion collisions. Regarding the
SU(3) gauge theory, $q=0.9835\pm0.0005$. Interestingly it is smaller
than $1$, although $|1-q|$ is in the same order of magnitude in both
models.

These results encourage us to proceed to our main goal, namely to
perform similar analysis for QCD and to compare the results with
experimental data.

This work was supported by the Hungarian Research
Fund (OTKA) under contract No. K104292.

\end{document}